\documentclass[aps,pra,reprint,superscriptaddress]{revtex4-1}

\usepackage{graphicx}
\usepackage{amsmath}
\usepackage{amssymb}
\usepackage[strict]{changepage}

\def\sp{@{\hspace{1.7mm}}}
\DeclareMathOperator{\var}{var}

\DeclareMathOperator{\Tr}{Tr}
\DeclareMathOperator{\re}{Re}

\begin{document}

\title{Fractional quantum Hall states in lattices: Local models and physical implementation}
\author{Anne E. B. Nielsen}
\affiliation{Max-Planck-Institut f{\"u}r Quantenoptik,
Hans-Kopfermann-Strasse 1, D-85748 Garching, Germany}
\author{Germ\'an Sierra}
\affiliation{Instituto de F\'isica Te\'orica, UAM-CSIC, Madrid, Spain}
\author{J. Ignacio Cirac}
\affiliation{Max-Planck-Institut f{\"u}r Quantenoptik,
Hans-Kopfermann-Strasse 1, D-85748 Garching, Germany}

\begin{abstract}
The fractional quantum Hall (FQH) effect is one of the most striking phenomena in condensed matter physics. It is described by a simple Laughlin wavefunction and has been thoroughly studied both theoretically and experimentally. In lattice systems, however, such an effect has not been observed, there are few simple models displaying it, and only few mechanisms leading to it are known. Here we propose a new way of constructing lattice Hamiltonians with local interactions and FQH like ground states. In particular, we obtain a spin 1/2 model with a bosonic Laughlin like ground state, displaying a variety of topological features. We also demonstrate how such a model naturally emerges out of a Fermi-Hubbard like model at half filling, in which the kinetic energy part possesses bands with nonzero Chern number, and we show how this model can be implemented in an optical lattice setup with present technology.
\end{abstract}

\maketitle

The FQH effect \cite{tsui} is one of the most fundamental phenomena in strongly correlated electronic systems and a paradigm of topological behaviour. It has been thoroughly studied and characterized in situations, where the solid surrounding the electrons has a modest effect on their properties \cite{yoshioka,stern,stormer}. This was possible, in part, thanks to the successful description in terms of the celebrated Laughlin wavefunctions \cite{laughlin1}. The appearance of FQH behaviour when the lattice structure created by the solid becomes important is, however, not so deeply studied nor understood. Nevertheless, the interest in lattice systems \cite{haldane1,wen,schroeter,greiter1,thomale,kapit,nielsen2,greiter2} has been sparked by the possibility of observing such behaviour with ultracold atoms \cite{sorensen,hafezi,bloch1,cooper}, obtaining high-temperature quantum Hall states \cite{tang,sun,neupert,roy,sheng,wang,regnault}, or even its applications in quantum information processing \cite{nayak}.

The standard route to the FQH effect in solids involves fractionally filled Landau levels (created by a strong, external magnetic field) and electron-electron interactions. This mechanism can be adapted to lattice systems by replacing the fractionally filled Landau level by a sufficiently flat and fractionally filled Chern band and adding near-neighbour interactions \cite{neupert,sheng,wang,regnault,parameswaran,yao}. A Chern band can be obtained by a proper choice of complex hopping amplitudes and can be made flat with longer range hoppings \cite{tang,sun,neupert}.

Here, we propose and investigate a different mechanism to obtain FQH states in lattice systems. We consider spin 1/2 particles on a square lattice with some specific short-range interactions. As we will show, the ground state of the system is extremely well described by the Kalmeyer-Laughlin (KL) wavefunction \cite{kalmeyer,laughlin2} (the spin version of the bosonic Laughlin wavefunction with Landau level filling factor $\nu=1/2$) and exhibits the topological behaviour expected for FQH states. In particular, we consider periodic boundary conditions and analyze the ground state (quasi) degeneracy, its response to twisting the boundary conditions, the local indistinguishability of the ground states, and the topological entanglement entropy. We also demonstrate how our model can be implemented with ultracold atoms in optical lattices. The temperatures, tunnelling amplitudes, and interactions required to observe the exotic topological behaviour are the same as the ones required to observe the Neel antiferromagnetic order in the standard Fermi-Hubbard model, so that our predictions can be tested with present or planned technology \cite{bloch,jordens,schneider,trotzky}.

The mechanism leading to FQH states can be viewed from two perspectives. The first takes a spin wavefunction with appropriate topological properties (like the KL state) that can be built out of correlators from a conformal field theory (CFT) (in our case the $SU(2)_1$ Wess-Zumino-Witten) and for which one can derive a (nonlocal) parent Hamiltonian. This Hamiltonian is then deformed into a local one without crossing a phase transition. The second perspective considers spin 1/2 fermions moving on a lattice with (not necessarily flat) Chern bands and strong local interactions. When the lattice filling is $1/2$ for both spin up and down, a Mott state is formed, and the spin state inherits the topological character of the Chern bands. The first view may be extended \cite{nielsen2,nielsen1} to other models to obtain, e.g., Moore-Read like states \cite{moore}. The second lends itself for physical implementations.

\section{Model}

\begin{figure*}
\resizebox{0.95\textwidth}{!}{\includegraphics{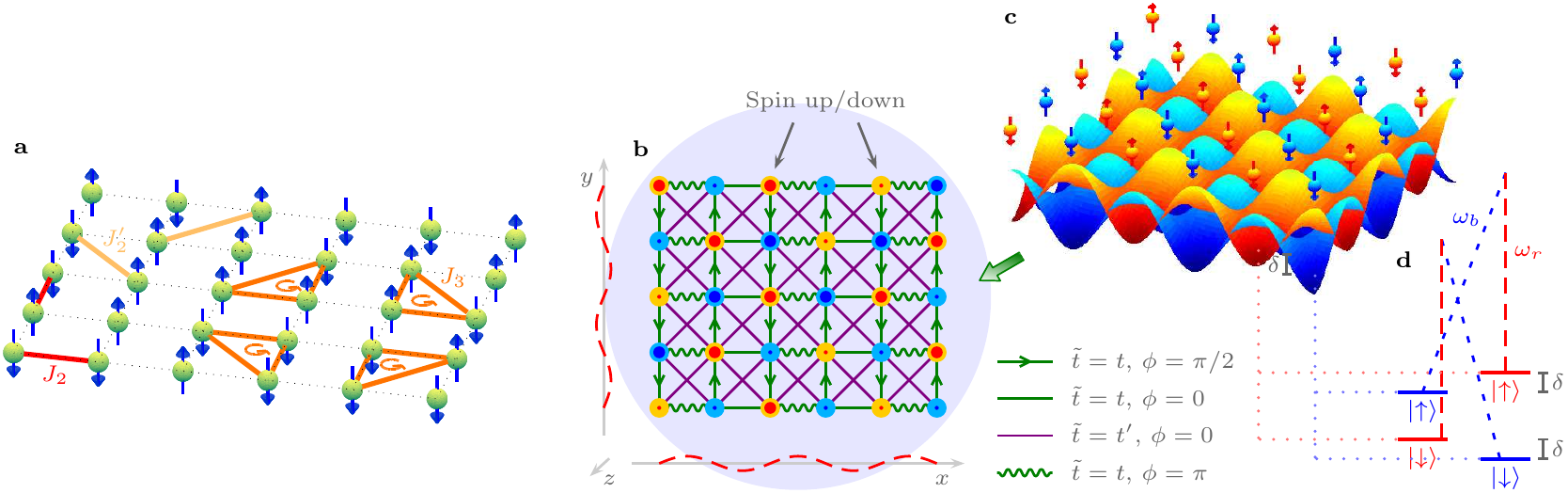}}
\caption{\textbf{Model and implementation.} \textbf{a}, The considered spin lattice Hamiltonian is a sum of local two- and three-body interactions. \textbf{b}, For suitable parameters, the spin lattice Hamiltonian is effectively equivalent to the Fermi-Hubbard like Hamiltonian in equation \eqref{FH}, of which we here show the kinetic energy parts $H_{\textrm{kin},\sigma}$, $\sigma\in\{{\uparrow},{\downarrow}\}$. Specifically, each arrow/line/wiggle from position $n$ to $m$ on the lattice represents the contribution
$\tilde{t}(e^{i\phi}\hat{a}_{m\sigma}^\dag\hat{a}_{n\sigma} +e^{-i\phi}\hat{a}_{n\sigma}^\dag\hat{a}_{m\sigma})$ to $H_{\textrm{kin},\sigma}$ with $\tilde{t}$ and $\phi$ given in the figure. \textbf{c}, $N$ fermions trapped in the optical lattice potential we propose to use for implementing the Fermi-Hubbard like Hamiltonian. \textbf{d}, We encode the spin up and down states in four internal hyperfine levels of the fermions. The blue/red states feel the blue/red potential in \textbf{c} and are hence trapped at the blue/red lattice sites. In this setting, we can implement the nearest-neighbour hopping terms through Raman transitions as indicated with the dashed blue and red lines in \textbf{d}. For this we need the three standing wave laser fields shown in \textbf{b} with electric fields: $\vec{E}_{r1}=\re(-2E\vec{\varepsilon}_z\sin(kx)e^{-i\omega_rT})$ (the dashed red line along the $x$-axis), $\vec{E}_{r2}=\re(-2iE\vec{\varepsilon}_z\sin(ky)e^{-i\omega_rT})$ (the dashed red line along the $y$-axis), and $\vec{E}_{b3}=\re(2(E_+\vec{\varepsilon}_++E_-\vec{\varepsilon}_-)\cos(k_zz)e^{-i\omega_bT})$ (the blue round shadow). Here, $E=E^*$ and $E_\pm$ are amplitudes, $\vec{\varepsilon}_z=(0,0,1)$, $\vec{\varepsilon}_\pm=(\mp1,-i,0)/\sqrt{2}$, $k=\omega_a/c=\pi/a$, where $c$ is the speed of light and $a$ is the lattice constant, $k_z=\omega_b/c$, $\omega_{r,b}$ are frequencies, and $T$ is time. The next-nearest neighbour hopping terms are implemented as a combination of hops induced by the fields listed above and tunnelling between nearest neighbour sites in the blue/red lattice.}\label{Fig1}
\end{figure*}

We start out with a spin wave function for an even number $N$ of spin 1/2 sitting at fixed positions in the two-dimensional plane
\begin{multline}\label{psiP}
\psi_\textrm{P0}^\textrm{CFT}(s_1,s_2,\ldots,s_N)=\\
\delta_\mathbf{s}\prod_{n=1}^N(-1)^{(n-1)(s_n+1)/2} \prod_{n<m}\left(z_n-z_m\right)^{(s_ns_m+1)/2}.
\end{multline}
Here, $z_n=x_n+iy_n$ is the position $(x_n,y_n)$ of the $n$th spin written as a complex number, $s_n=\pm1$ labels the possible states (`up' and `down') of the $n$th spin, and $\delta_\mathbf{s}=1$ for $\sum_ns_n=0$ and $\delta_\mathbf{s}=0$ otherwise. \eqref{psiP} is closely related to the KL state and reduces exactly to it in the case of a square lattice of infinite extent \cite{nielsen2}. We shall refer to \eqref{psiP} as the `CFT state' because it can be expressed as a CFT correlator (see the supplementary information).

An exact parent Hamiltonian of $\psi_\textrm{P0}^\textrm{CFT}$ has been derived in \cite{nielsen2} by use of a quite general technique relying on CFT. The Hamiltonian consists of interactions between all pairs and triples of spins in the system, which is challenging to achieve experimentally. One may argue, however, that deforming the Hamiltonian will not dramatically change the physical properties of the ground state as long as no phase transition is crossed. With this in mind and specializing to an $L_x\times L_y$ square lattice with $L_x$ even, we investigate the local Hamiltonian that remains after removing all long-range interactions and making all coupling strengths position independent:
\begin{multline}\label{Ham}
H=J_2\!\!\!\!\sum_{<n,m>}\!\!\! 2\vec{S}_n\cdot\vec{S}_m
+J'_2\!\!\!\!\!\sum_{<\!\!<n,m>\!\!>}\!\!\!\!\! 2\vec{S}_n\cdot\vec{S}_m\\
-J_3\!\!\!\!\!\!\!\sum_{<n,m,p>_{\circlearrowleft}}\!\!\!\!\!\!
4\vec{S}_n\cdot\left(\vec{S}_m\times\vec{S}_p\right).
\end{multline}
Here, $\vec{S}_n=(S_n^x,S_n^y,S_n^z)$ is the spin operator of the $n$th spin. As indicated in figure \ref{Fig1}a, the first (second) two-body term is summed over all pairs of nearest (next-nearest) neighbour spins, and the three-body term, which breaks time reversal symmetry, is summed over all triangles of neighbouring spins (for each triangle only one term is included and $n,m,p$ label the vertices of the triangle in the counter clockwise direction as indicated with the arrow). In this and the following section, we choose $J_2=1$, $J'_2=0$, and $J_3=1/2$ to make $H$ as local as possible. The Hamiltonian conserves the total spin $\vec{S}_\textrm{tot}=\sum_n\vec{S}_n$ and is hence $SU(2)$ invariant.

\begin{table}
\begin{tabular}{c \sp|\sp c \sp|\sp c \sp|\sp c \sp c \sp c}
\multicolumn{1}{c \sp}{} & & Plane & \multicolumn{2}{c\sp}{Torus} \\
N & $L_x\times L_y$ & $|\langle\psi_{\textrm{P}0}|\psi_{\textrm{P}0}^\textrm{CFT}\rangle|$ &
$|\langle\psi_{\textrm{T}0}|\psi_{\textrm{T}0}^{\textrm{CFT}}\rangle|$ & $|\langle\psi_{\textrm{T}1}|\psi_{\textrm{T}1}^{\textrm{CFT}}\rangle|$\\
\hline
12 & $4\times3$ & 0.9860 & 0.9818 & 0.9533\\
16 & $4\times4$ & 0.9812 & 0.9747 & 0.9572\\
20 & $4\times5$ & 0.9728 & 0.9655 & 0.9200\\
30 & $6\times5$ & - & 0.9258 & 0.9361\\
\end{tabular}
\caption{\textbf{Overlap between the ground state(s) of $H$ and the CFT state(s).}
Columns 1-2: We consider $N$ spins on a square lattice of size $L_x\times L_y$. Column 3: The overlap between the ground state $\psi_{\textrm{P}0}$ of the local Hamiltonian in equation \eqref{Ham} with open boundary conditions and the CFT state in equation \eqref{psiP}. Columns 4-5: On the torus, i.e.\ for periodic boundary conditions, there are two CFT states (see supplementary equations (S9-S10)), and the columns show the overlap of these states with the two lowest energy eigenstates  $\psi_{\textrm{T}0}$ and $\psi_{\textrm{T}1}$ of \eqref{Ham}. The overlaps are remarkably high, in particular when taking the large dimension of the involved Hilbert spaces into account (e.g., $1.1\times10^9$ for a $6\times5$ lattice). We include fewer results for the plane than for the torus because the lack of translational invariance in the plane makes it harder to diagonalize the Hamiltonian in this geometry.}\label{Tab1}
\end{table}

To test how much the deformation of the Hamiltonian affects the ground state, we compute the overlap between the ground states before and after the deformation in table \ref{Tab1} column 3. For many-body systems, the dimension of the Hilbert space grows exponentially with the number of spins in the system, and the hugeness of the Hilbert space may easily cause states in the same phase to have poor overlaps. The fact that we obtain high overlaps for states with a few tenth of spins, on the other hand, is a strong indication that the deformation of the Hamiltonian does not bring the system to a different phase. In fact, we may regard \eqref{psiP} as an analytical approximation to the ground state, which is extremely helpful to see the underlying physics and to do computations for systems that are too large for exact diagonalization. We have chosen the parameters of the Hamiltonian to get large overlaps, but we note that high overlaps are obtained in a broad parameter region as we show below.

The model we have looked at so far is defined in the plane and hence has open boundary conditions. To investigate the topological properties of $H$ and the CFT state, we shall also need to consider periodic boundary conditions, in which case $H$ is translationally invariant. On the torus, i.e., for periodic boundary conditions in both directions, there are two CFT states, $\psi^{\textrm{CFT}}_{\textrm{T}0}$ and $\psi^{\textrm{CFT}}_{\textrm{T}1}$, for which we derive explicit analytical expressions in the supplementary information. Rewriting these expressions for the case of a square lattice, we find that the CFT states are exactly the KL states on the torus for all $L_x$ and $L_y$. Table \ref{Tab1} shows that the large overlap between the ground states of $H$ and the CFT states also hold on the torus.

\section{Topological properties}

FQH states are examples of topological states, and in the following we demonstrate that $H$ and the CFT states exhibit topological properties that fit the properties of the $\nu=1/2$ Laughlin state in the continuum. Characteristic features of topological states are a ground state degeneracy that depends on the topology of the surface on which the states are defined, the lack of ability to distinguish the topologically degenerate states through local measurements, and a nonzero topological entanglement entropy. For FQH states in solids another important quantity is the Hall conductivity. The Hall conductivity is closely related to the Chern number \cite{thouless1,niu,kohmoto}, which can also be computed for spin systems \cite{hatsugai}. For Laughlin states the expected behaviour is a many-body Chern number of the ground state manifold on the torus that is unity and spectral flow of the ground states into each other for certain choices of the lattice size under continuous twisting of the boundary conditions \cite{regnault,thouless2}.

\begin{figure}
\resizebox{0.75\columnwidth}{!}{\includegraphics{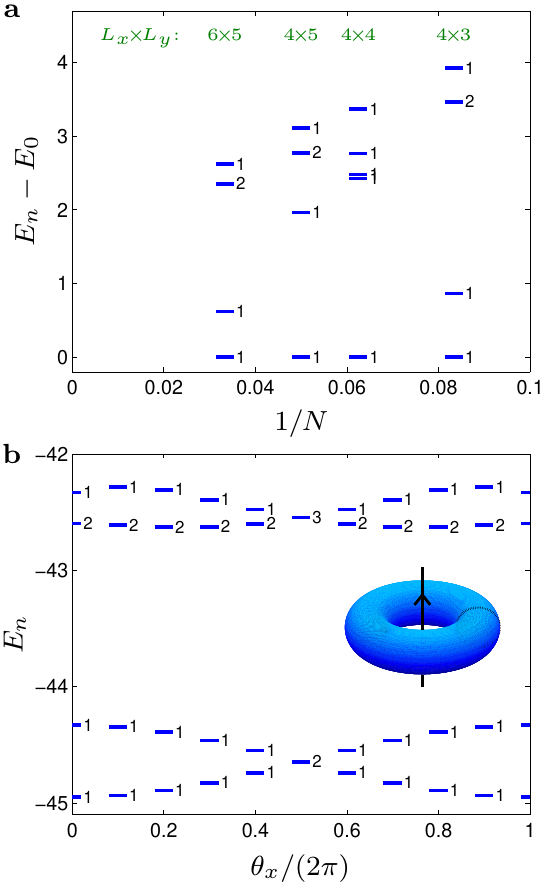}}
\caption{\textbf{Topological properties of the Hamiltonian.} \textbf{a}, Energy spectrum in the $S_\textrm{tot}^z=0$ subspace on the torus for different lattice sizes (in green, $N=L_xL_y$) obtained by exact diagonalization. The number to the right of each energy level is the degeneracy and only the energies of the five lowest states ($n=0,1,2,3,4$) are displayed. Note that the spectrum in the complete Hilbert space is the same except that each state is replaced by $2S+1$ degenerate states, where $S$ is the total spin quantum number of the state. As the two lowest states have $S=0$, the results suggest that there are two degenerate ground states and a gap to the first excited state in the thermodynamic limit like for the $\nu=1/2$ Laughlin state in the continuum \cite{haldane2}. \textbf{b}, Energies of the five lowest states in the $S_\textrm{tot}^z=0$ subspace on the torus for twisted boundary conditions in the $x$-direction ($\theta_x$ is the twist angle) and a lattice size of $6\times5$. Twisting the boundary conditions corresponds to gradually inserting a flux line through the hole of the torus (inset), and we observe that the two ground states flow into each other under this operation.}\label{Fig2}
\end{figure}

We compute the spectrum on the torus through exact diagonalization \cite{laeuchli} of $H$ (figure \ref{Fig2}a). The exponential increase of the dimension of the Hilbert space with the number of spins limits the system sizes we can consider. It is seen that the two lowest energy states approach each other as the system size is increased, and extrapolating the energy difference between the third lowest and the lowest energy state to the limit $N\rightarrow\infty$ points to the existence of a gap in the thermodynamic limit.

\begin{figure}
\resizebox{0.8\columnwidth}{!}{\includegraphics{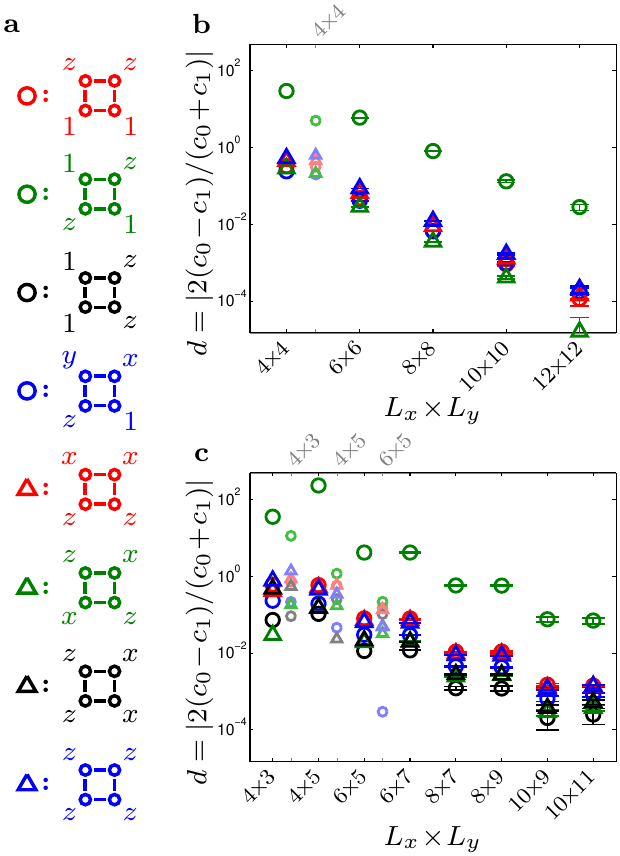}}
\caption{\textbf{Indistinguishability of local observables.} \textbf{a}, To demonstrate local indistinguishability of the states on the torus, we consider the set of all spin operators that act on a plaquette of four spins. Note that all plaquettes are equivalent due to the translational invariance. Using symmetries and the properties of spin operators, the correlators of all such local operators can be expressed in terms of the eight correlators depicted (the uppermost drawing, e.g., represents the correlator $c(\psi)=\langle\psi|S_{n_1}^zS_{n_2}^z|\psi\rangle/\langle\psi|\psi\rangle$, where the spins on the plaquette are labelled $n_1$, $n_2$, $n_3$, $n_4$ in the counter clockwise direction starting from the upper right corner). For the special case $L_x=L_y$, the correlators displayed in black are not needed. \textbf{b} and \textbf{c}, Dependence of the relative difference $d=|2(c_0-c_1)/(c_0+c_1)|$ between the correlators $c_0\equiv c(\psi_{\textrm{T}0}^{\textrm{CFT}})$ and $c_1\equiv c(\psi_{\textrm{T}1}^{\textrm{CFT}})$ on the size of the lattice for even-by-even and even-by-odd lattices (we use the markers indicated in panel \textbf{a}). The extra set of smaller fainter symbols for $4\times4$ in \textbf{b} and $4\times3$, $4\times5$, and $6\times5$ in \textbf{c} (see the upper axes) show the same for the exact ground states of the Hamiltonian. The results are obtained by exact computations for $L_xL_y\leq30$ and from Monte Carlo simulations for $L_xL_y\geq36$. The error bars of $d$ are given as $d\pm\delta d$, where $\delta d=\sqrt{(\frac{\partial d}{\partial c_0})^2(\delta c_0)^2+(\frac{\partial d}{\partial c_1})^2(\delta c_1)^2}$, $\delta c_i=\sqrt{\var{(c_i)}/\mathcal{N}}$, and the variance is taken over the outcome of $\mathcal{N}$ independent Monte Carlo trajectories.}\label{Fig3}
\end{figure}

For finite size systems, the local indistinguishability of topologically degenerate states is not perfect, but the ability to distinguish the states decreases exponentially with the size of the system. In figure \ref{Fig3}, we plot the deviation $d$ between the correlators of local operators computed for the two states  on the torus for both the exact ground states and the CFT states. In both cases, $d$ decreases with increasing system size, and for the CFT states the decay is clearly exponential. We also observe that, as far as the operators appearing in the Hamiltonian are concerned, $d$ is not small unless the lattice size is at least $6\times5$. This explains why the ground state degeneracy first appears approximately for the $6\times5$ lattice in figure \ref{Fig2}a, and the general decrease of $d$ in figure \ref{Fig3} provides further evidence that the two lowest energies will approach each other further for larger lattices.

\begin{figure}
\resizebox{0.75\columnwidth}{!}{\includegraphics{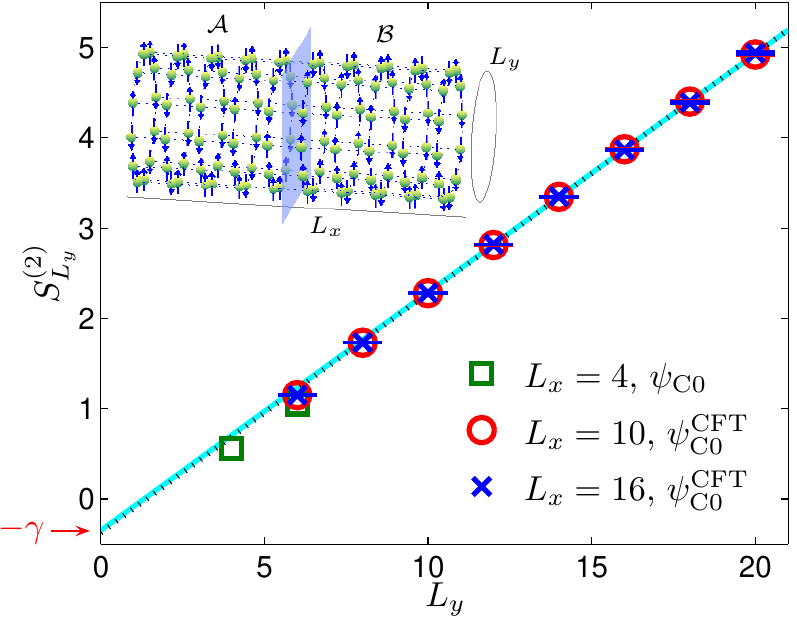}}
\caption{\textbf{Topological entanglement entropy.} To compute the topological entanglement entropy, we map the state in equation \eqref{psiP} to a cylinder (inset) with closed ends at $\pm\infty$ (see supplementary information for details). We then cut the cylinder in two halves (part $\mathcal{A}$ and $\mathcal{B}$) as indicated with the blue plane, and compute the entanglement entropy $S_{L_y}^{(2)}\equiv-\ln(\Tr(\rho_\mathcal{A}^2))$ as a function of $L_y$ for fixed $L_x$, where $\rho_{\mathcal{A}}$ is the reduced density operator of part $\mathcal{A}$. We choose here to use the Renyi entropy with index 2 because it is less demanding to compute numerically than the von Neumann entropy \cite{hastings}. For large $L_x$ and $L_y$, $S_{L_y}^{(2)}$ is independent of $L_x$ and grows linearly with $L_y$, and the intersection with the vertical axis is the topological entanglement entropy ($-\gamma$) \cite{jiang,kitaev,levin,haque,zhang}. The fact that the results for $L_x=10$ and $L_x=16$ practically coincide shows that we are already in the $L_x\rightarrow\infty$ limit. The black dotted line is a linear fit to the results for $L_x=10$ and $L_y\geq10$ and gives $\gamma=0.374$. The solid cyan line is the same fit except that $\gamma$ is fixed to the value $\gamma=\ln(2)/2\approx0.347$ of the $\nu=1/2$ Laughlin state, and it is seen that both fits fit the data well. The results for $L_x=4$ are computed using the lowest energy state of the Hamiltonian in equation \eqref{Ham} with periodic boundary conditions in the $y$-direction, and they approximately follow the results for the CFT state. The results for $L_x=10$ and $L_x=16$ are computed from Monte Carlo simulations of $r\equiv\exp(-S_{L_y}^{(2)})$. The error bars are given as $S_{L_y}^{(2)}\pm\delta S_{L_y}^{(2)}$, where $\delta S_{L_y}^{(2)}=\left|\frac{\partial \ln(r)}{\partial r}\right|\sqrt{\var(r)/\mathcal{N}}$ and the variance is taken over the outcome of $\mathcal{N}$ independent Monte Carlo trajectories.}\label{Fig4}
\end{figure}

In figure \ref{Fig4}, we compute the topological entanglement entropy $-\gamma$ using the approach proposed in \cite{jiang}. For the $\nu=1/2$ Laughlin state in the continuum, $\gamma=\ln(2)/2$ \cite{fendley}, and the result we get for the CFT state is in perfect agreement with this value. For the exact ground state of $H$, we can again only consider rather small systems, but we observe that the results approximately follow those of the CFT state and also point to a nonzero value of $\gamma$.

We have computed the many-body Chern number for both the exact ground states and the CFT states on the torus for a $4\times5$ lattice using the method described in \cite{hafezi,hatsugai}, and in both cases we get $1$. Figure \ref{Fig2}b demonstrates the flow of the ground states of $H$ into each other when the boundary conditions are twisted, and for the CFT states this can be shown analytically. Again, our findings are thus in agreement with the properties of the $\nu=1/2$ Laughlin state.

\section{Connection to a Fermi-Hubbard like model}

As mentioned in the introduction, each spin in our model may represent the spin of a fermion sitting on a site in an optical lattice. Let us consider the Fermi-Hubbard like Hamiltonian
\begin{equation}\label{FH}
H_{\textrm{FH}}=\sum_\sigma H_{\textrm{kin},\sigma} +U\sum_{n=1}^N\hat{a}_{n{\uparrow}}^\dag \hat{a}_{n{\uparrow}}
\hat{a}_{n{\downarrow}}^\dag \hat{a}_{n{\downarrow}}
\end{equation}
on a square lattice, where $\hat{a}_{n\sigma}$ annihilates a fermion with spin $\sigma$ on lattice site $n$, $U$ is a positive constant, $H_{\textrm{kin},\sigma}$ is defined in figure \ref{Fig1}b, and we choose the total number of fermions to equal the number of lattice sites. $H_{\textrm{kin},\sigma}$ represents spin preserving hopping of fermions between nearest and next-nearest neighbouring sites with complex hopping amplitudes, and the second term in $H_{\textrm{FH}}$ represents interactions between fermions sitting on the same site.

When $U$ is much larger than the hopping strengths $t,t'$, we are in the Mott insulating regime, where each site is occupied by a single fermion. In this limit, we can use the Schrieffer-Wolff transformation \cite{bravyi} to derive an effective Hamiltonian $H_{\textrm{eff}}$ acting in the space of states with only single occupancy on all sites in the same way as the Heisenberg model is derived from the standard Fermi-Hubbard model. Reexpressing $H_{\textrm{eff}}$ in terms of spins, we get $H_{\textrm{eff}}=H+\textrm{constant}$ to third order in $t/U$, where $H$ is given by equation \eqref{Ham} with $J_2=2t^2/U$, $J'_2=2t'^2/U$, and $J_3=6t^2t'/U^2$. This is not by chance: the $SU(2)$ invariance of $H_\textrm{FH}$ is automatically inherited to $H_\textrm{eff}$ and the chirality is built into the model through the complex hopping amplitudes.

\begin{figure}
\resizebox{\columnwidth}{!}{\includegraphics{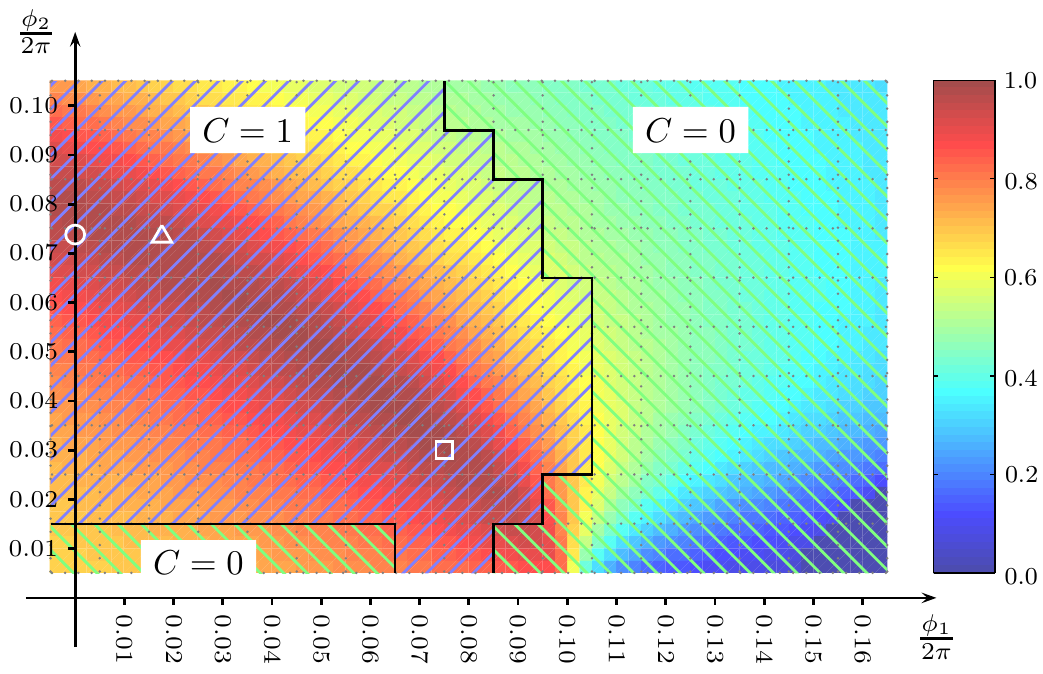}}
\caption{\textbf{Phase diagram.} Phase diagram of the Hamiltonian in equation \eqref{Ham} on the torus with $J_2=\cos(\phi_1)\cos(\phi_2)$, $J'_2=\sin(\phi_1)\cos(\phi_2)$, and $J_3=\sin(\phi_2)$. $C$ is the total Chern number of the states $\psi'_{\textrm{T}0}$ and $\psi'_{\textrm{T}1}$ on a $4\times5$ lattice, where $\psi'_{\textrm{T}0}$ ($\psi'_{\textrm{T}1}$) is the lowest energy state in the subspace spanned by all states with the same eigenvalues of $S_\textrm{tot}^z$ and the translation operators in the $x$- and $y$-directions as $\psi_{\textrm{T}0}^\textrm{CFT}$ ($\psi_{\textrm{T}1}^\textrm{CFT}$). Within the topological phase ($C=1$), the two states are well separated from higher energy states in the same subspaces and flow into each other under flux insertion (like in figure \ref{Fig2}\textbf{b}). The background colour gives the overlap between the CFT state \eqref{psiP} and the ground state of \eqref{Ham} for a $4\times5$ lattice with open boundary conditions. The white circle marks the Hamiltonian studied in the first part of the article and the white triangle and the white square mark possible parameter choices discussed in the text. We omit $\phi_2=0$ because the additional symmetries present for this case may cause the lowest energy states in the considered subspaces to be degenerate.}\label{Fig5}
\end{figure}

The phase diagram of $H$ on the torus based on Chern number computations for a $4\times5$ lattice is depicted in figure \ref{Fig5}. As the $4\times5$ lattice is too small to display a clear two-fold ground state degeneracy, it may happen that the two lowest energy states are not the states that resemble the CFT states. We circumvent this problem by noting that the Hamiltonian is block diagonal, which allows us to select the right states based on quantum numbers as stated in the caption. The diagram displays the freedom we have to vary the coupling strengths in the Hamiltonian while staying within the topological phase. In particular, it reveals a ridge with high overlap between the CFT state and the ground state of $H$ that connects the Hamiltonian studied in the previous sections with the region where $t,t'\ll U$. Choosing, e.g., $\phi_1=0.075\times2\pi$ and $\phi_2=0.03\times2\pi$ (the white square in the figure), we get $t/U=0.10$, $t'/U=0.07$, and an overlap of $0.985$.

The observation that $H$ and $H_\textrm{FH}$ are closely related provides some insight into how our model compares to previous proposals. Specifically, $H_\textrm{FH}$ can be seen as a sum of two free fermion models, $H_{\textrm{kin},{\uparrow}}$ and $H_{\textrm{kin},{\downarrow}}$, and a local interaction term. Each of the free models has two bands of which the lowest is filled and the highest is empty. The Chern number of the filled band (which we compute using equation (12) in \cite{shi}) is plus or minus one for all nonzero $t$ and $t'$, and the band flattening, i.e., the ratio of the gap between the bands to the width of one of the bands, is moderate $F\leq2/(\sqrt{2}-1)\approx4.8$. Our results thus show that two copies of a fermionic Chern insulator with integer band filling plus local interactions can give rise to a bosonic FQH state. This behaviour is different from flat band models, where a fermionic (bosonic) Chern insulator with flattened and fractionally filled bands plus interactions give rise to a fermionic (bosonic) FQH state \cite{neupert,sheng,wang}. We therefore conclude that our model provides a different mechanism to get the FQH effect in lattices.

\section{Simulation scheme}

The Fermi-Hubbard like model can be simulated in fermions in optical lattices as follows. We encode the spin up and down states in four internal hyperfine levels as shown in figure \ref{Fig1}d. By choosing the internal states appropriately and using polarized light, it is possible to construct the optical lattice in figure \ref{Fig1}c, where the levels shown in blue (red) are trapped in a potential with minima located at the white (black) squares of a checkerboard. As detailed in the supplementary information, this can be done with eight laser beams travelling in the directions $(\pm1,\pm1,\pm\beta)$, where $\beta$ is adjustable.

In this setup, the atoms can tunnel through the potential barrier between nearest neighbour sites in the blue/red lattice, which provides a contribution to the next-nearest neighbour hopping terms in the Hamiltonian. We propose to implement the nearest neighbour hopping terms by use of laser assisted tunnelling \cite{jaksch}. This can be done with the three standing wave laser fields in figure \ref{Fig1}b, which we refer to as r1, r2, and b3, respectively. As explained further in the supplementary information, r1 and b3 (r2 and b3) induce nearest neighbour hops in the horizontal (vertical) direction. A hop is accompanied by a Raman transition between internal states as illustrated in figure \ref{Fig1}d. The alternating signs of the hopping amplitudes originate from the spatial oscillations of the amplitudes of r1 and r2 with a period that is twice the lattice spacing, and the $i$'s on the hopping amplitudes in the vertical direction appear because the amplitude of r2 is imaginary. Note that it is possible to choose the internal states in such a way that the energy difference between the blue and red spin up states is the same as the energy difference between the blue and red spin down states. r1, r2, and b3 therefore accomplish the hops for both spin up and spin down.

A particularly convenient feature of the above model and implementation scheme is that only on-site interactions are needed. These occur naturally in the optical lattice, and so the Hamiltonian in equation \eqref{FH} is implemented directly. The Hamiltonian \eqref{Ham} captures the physics of \eqref{FH} when $t,t'\ll U$ and the temperature is small compared to the gap to the first excited state, which is of order $t^2/U$. The requirements regarding temperature, interactions, and hopping amplitudes are thus the same as those needed to observe the Neel antiferromagnetic order in the standard Fermi-Hubbard model, which several groups are pursuing at the moment. Even if \eqref{FH} does not exactly reproduce \eqref{Ham}, it is possible that similar physics can be observed. This suggests that, e.g., the choice $t/U=1/2$ and $t'/U=1/6$ (corresponding to the white triangle in figure \ref{Fig5}), which leads to more moderate temperature requirements, may already give interesting results.

\section*{Acknowledgements}

The authors acknowledge discussions with Fabio Mezzacapo, Hong-Hao Tu, and Shuo Yang. This work has been supported by the EU project AQUTE, FIS2012-33642, QUITEMAD, and the Severo Ochoa Program.

\section*{Author contributions}

All authors planned the project together and contributed to the ideas behind the work, AEBN and GS did the analytical computations related to the CFT states, AEBN did all other analytical and numerical computations and drafted the manuscript, AEBN and JIC developed the simulation protocol, and all authors contributed to the final version of the manuscript.

\section*{Competing financial interests}

The authors declare no competing financial interests.

\onecolumngrid
\setcounter{equation}{0}
\setcounter{figure}{0}
\newpage

\newcommand{\newsec}[1]{%
\vspace{6mm}
\noindent
{\textbf{#1}}
\vspace{1mm}}

\changepage{}{-6mm}{}{0mm}{}{}{}{}{}

\renewcommand{\theequation}{S\arabic{equation}}
\renewcommand{\thefigure}{S\arabic{figure}}

\begin{center}
{\large \textit{Supplementary information}}
\end{center}

\section*{CFT states from the $SU(2)_1$ Wess-Zumino-Witten model}

\vspace{-5mm}

\newsec{Background}

\noindent The Wess-Zumino-Witten (WZW) model based on the Kac-Moody group $SU(2)_1$ is a CFT with central charge $c=1$ and two primary fields $\phi_0(z)$ and $\phi_{1/2}(z)$ with conformal weights $h_0=0$ and $h_{1/2}=1/4$, respectively. The field $\phi_j(z)$ is associated with the total spin $j$ and has components $\phi_{j,m}(z)$ with $m=-j,-j+1,\ldots,j$. From these fields one can construct chiral correlators, also called conformal blocks,
\begin{equation}\label{cb}
\tilde{\psi}_k^{\textrm{CFT}}(m_1,m_2,\ldots,m_N)= \langle\phi_{j_1,m_1}(z_1)\phi_{j_2,m_2}(z_2)\cdots\phi_{j_N,m_N}(z_N)\rangle_k,
\end{equation}
and interpret the result as a quantum state $|\tilde{\psi}_k^\textrm{CFT}\rangle= \sum_{m_1,m_2,\ldots,m_N}\tilde{\psi}_k^{\textrm{CFT}} (m_1,m_2,\ldots,m_N)|m_1,m_2,\ldots,m_N\rangle$ of $N$ particles with spins $j_n$ at fixed positions $z_n$. $\langle\,\bullet\,\rangle$ stands for vacuum expectation value, the word chiral refers to the fact that only $z_n$ and not $\bar{z}_n$ appears on the right hand side of \eqref{cb}, and the index $k$ takes into account that there may be more than one conformal block for a given choice of fields. The number of conformal blocks on a given Riemann surface with genus $g$ is dictated by the fusion rules of the primary fields. Specifically, for $N$ primary fields $\phi_{1/2}$ (which is the interesting case since $\phi_0$ represents the identity) the number of conformal blocks is $2^g$ if $N$ is even, whereas the chiral correlator vanishes if $N$ is odd. Hence, on the sphere, i.e.\ $g=0$, there is only one chiral correlator, while on the torus, i.e.\ $g=1$, there are two.

\newsec{CFT state in the plane}

\noindent The plane geometry considered in the main text is the Riemann sphere. The chiral correlator is well-known in this geometry, so here we simply note that the CFT expression leading to equation (1) in the main text is
\begin{equation}
\psi_\textrm{P0}^\textrm{CFT}(s_1,s_2,\ldots,s_N)\propto
\langle\phi_{\frac{1}{2},\frac{s_1}{2}}(z_1) \phi_{\frac{1}{2},\frac{s_2}{2}}(z_2) \ldots \phi_{\frac{1}{2},\frac{s_N}{2}}(z_N)\rangle
\end{equation}
with $N$ even.

\newsec{CFT state on the cylinder with closed ends}

\noindent The plane (or more precisely Riemann sphere) can be mapped to the cylinder with closed ends and circumference $L_y$ via the coordinate transformation $z=e^{2\pi(l_x+il_y)/L_y}$, where $z$ is a point in the plane written as a complex number and $(l_x,l_y)$ is the corresponding point on the cylinder (note that $l_y$ is periodic with period $L_y$, whereas $l_x$ is not). The cylinder has closed ends, and hence genus zero, because $e^{2\pi (l_x+il_y)/L_y}$ is a single point rather than a circle for $l_x=\pm\infty$. The wavefunction $\psi_{\textrm{C}0}^\textrm{CFT}$ on the cylinder is therefore easily obtained from $\psi_\textrm{P0}^\textrm{CFT}$ by choosing $z_n=e^{2\pi (l_{x,n}+il_{y,n})/L_y}$ for $n=1,2,\ldots,N$. For the particular case of a square lattice, we choose $l_{x,n}\in\{-(L_x-1)/2,-(L_x-1)/2+1,\ldots,(L_x-1)/2\}$ and $l_{y,n}\in\{1,2,\ldots,L_y\}$.

\newsec{CFT states on the torus}

\noindent To derive the conformal blocks on the torus, we use the fact that the $SU(2)_1$ WZW model can be bosonized in terms of a massless free scalar field $\varphi(z,\bar{z})$ compactified on a circle of radius $R=\sqrt{2}$. This leads to the result
\begin{equation}\label{ver}
\left\langle \prod_{i=1}^N : e^{ i \frac{s_i}{\sqrt{2}}  \varphi(z_i) + i \frac{s_i}{\sqrt{2}} \bar{\varphi} (\bar{z_i})  }  : \right\rangle=\sum_k\left\langle\phi_{\frac{1}{2},\frac{s_1}{2}}(z_1) \phi_{\frac{1}{2},\frac{s_2}{2}}(z_2) \ldots \phi_{\frac{1}{2},\frac{s_N}{2}}(z_N)\right\rangle_k
\left\langle\bar{\phi}_{\frac{1}{2},\frac{s_1}{2}}(\bar{z}_1) \bar{\phi}_{\frac{1}{2},\frac{s_2}{2}}(\bar{z}_2) \ldots \bar{\phi}_{\frac{1}{2},\frac{s_N}{2}}(\bar{z}_N)\right\rangle_k,
\end{equation}
where $\varphi(z_i)$ ($\bar{\varphi}(\bar{z_i})$) is the chiral (antichiral) component of the scalar field $\varphi(z,\bar{z})=\varphi(z)+\bar{\varphi}(\bar{z})$ and $:\bullet:$ denotes normal ordering. An expression for the correlator on the left hand side has been obtained in \cite{S_dijkgraaf}, and by rewriting this expression we can extract the conformal blocks \eqref{cb} up to phase factors that are independent of $z_1,z_2,\ldots,z_N$. The phase factors are determined by noting that the conformal blocks on the torus reduce to the conformal block on the plane if all the spins sit within a small region.

Before writing down the result for the conformal blocks, we note that a torus can be defined by specifying two complex numbers $\omega_1$ and $\omega_2$ and identifying points in the complex plane that differ by $n\omega_1+m\omega_2$, where $n$ and $m$ are integers. (The appropriate choice for the square lattice, e.g., is $\omega_1=L_x$ and $\omega_2=iL_y$.) The modular parameter $\tau=\omega_2/\omega_1$ is assumed to have positive imaginary part without loss of generality. Defining the scaled positions $w_n=z_n/\omega_1$, the conformal blocks read
\begin{equation}\label{cbt}
\tilde{\psi}^{\textrm{CFT}}_k(s_1,s_2,\ldots,s_N) =\frac{\delta_\mathbf{s}}{\eta(q)}(-1)^{\sum_{n=1}^N(n-1)(s_n+1)/2} \;
\theta_{k,0}\left(\sum_{n=1}^N w_ns_n,2\tau\right)
\prod_{n<m}E(w_n,w_m)^{s_ns_m/2},
\end{equation}
where $k$ takes the values $0$ and $1/2$,
\begin{equation}
\delta_\mathbf{s}=\left\{\begin{array}{ll}
1 & \textrm{for } \sum_ns_n=1 \\
0 & \textrm{otherwise}
\end{array}\right.,
\end{equation}
$\eta(q)$ is the Dedekin eta function,
\begin{equation}
\eta(q)=q^{1/24}\prod_{n=1}^\infty(1-q^n), \qquad q=e^{2\pi i\tau},
\end{equation}
$\theta_{a,b}(w,\tau)$ is the theta function,
\begin{equation}
\theta_{a,b}(w,\tau)=\sum_{n=-\infty}^{\infty}e^{i\pi\tau(n+a)^2+2\pi i(n+a)(w+b)},
\end{equation}
$E(w_n,w_m)$ is the prime form on the torus,
\begin{equation}
E(w_n,w_m)=\frac{\theta_{\frac{1}{2},\frac{1}{2}}(w_n- w_m,\tau)}{\theta_{\frac{1}{2},\frac{1}{2}}'(0,\tau)}, \qquad \theta_{\frac{1}{2},\frac{1}{2}}'(0,\tau)= \left.\frac{d\theta_{\frac{1}{2},\frac{1}{2}}(w, \tau)}{dw}\right|_{w=0},
\end{equation}
and the operation $(\bullet)^{s_ns_m/2}$ is made single valued by always choosing the phase of $E(w_n,w_m)$ to be within the interval $]-\pi,\pi]$.

Let us consider the special case of an $L_x\times L_y$ square lattice and choose the origin such that $\sum_nw_n=0$. If we regard a spin up as an occupied site and a spin down as an empty site, we find after several analytical manipulations that \eqref{cbt} is proportional to the corresponding continuum FQH states provided in \cite{S_haldane,S_read} except that the possible positions of the particles are restricted to the lattice and except for some sign factors that make the state into a spin singlet. In other words, the CFT states are precisely the Kalmeyer-Laughlin states on the torus. We note also that the states in \cite{S_read} allow for twisted boundary conditions and therefore provide a natural way to generalize the CFT states given below to this case.

The CFT states on the torus are obtained as linear combinations of the conformal blocks \eqref{cbt} chosen in such a way that the CFT states reflect the symmetries of the Hamiltonian in equation (2) in the main text. Specifically, we would like the CFT states to be eigenstates of the translation operators, $T_x$ and $T_y$, in the $x$- and $y$-directions. Acting with $T_x$ and $T_y$ on the conformal blocks, we find after a lengthy computation that the appropriate choice for an even-by-odd lattice is
\begin{equation}
\begin{array}{l}
\psi_{\textrm{T}0}^\textrm{CFT}(s_1,s_2,\ldots,s_N) \propto\tilde{\psi}^{\textrm{CFT}}_0(s_1,s_2,\ldots,s_N)\\[2mm]
\psi_{\textrm{T}1}^\textrm{CFT}(s_1,s_2,\ldots,s_N) \propto\tilde{\psi}^\textrm{CFT}_{1/2}(s_1,s_2,\ldots,s_N)
\end{array}
\qquad \textrm{(even-by-odd lattice)}.
\end{equation}
For the even-by-even lattice with $L_x=L_y$ both of the conformal blocks are invariant under translations, so we need to consider additional symmetries. Requiring the CFT states to be eigenstates of the operator that rotates the lattice by ninety degrees, we get
\begin{equation}
\begin{array}{l}
\psi_{\textrm{T}0}^\textrm{CFT}(s_1,s_2,\ldots,s_N) \propto\tilde{\psi}^{\textrm{CFT}}_0(s_1,s_2,\ldots,s_N)
+\left(-1+\sqrt{2}\right)\tilde{\psi}^{\textrm{CFT}}_{1/2}(s_1,s_2,\ldots,s_N)\\[2mm]
\psi_{\textrm{T}1}^\textrm{CFT}(s_1,s_2,\ldots,s_N) \propto\tilde{\psi}^\textrm{CFT}_0(s_1,s_2,\ldots,s_N)
+\left(-1-\sqrt{2}\right)\tilde{\psi}^\textrm{CFT}_{1/2}(s_1,s_2,\ldots,s_N)
\end{array}
\qquad (L_x=L_y).
\end{equation}
This construction also guarantees the CFT states to be orthogonal, which is not necessarily the case for the conformal blocks.

\section*{Optical lattice potential}

The main idea behind the scheme we propose to realize the optical lattice potential in figure 1c in the main text is to encode the blue (red) states in internal levels that interact more strongly with right (left) circularly polarized light than with left (right) circularly polarized light and then create a checkerboard pattern of regions with highest intensity of right circularly polarized light and regions with highest intensity of left circularly polarized light. Since the blue and red sublattices are square lattices when viewed from the $(x+y)$- or $(x-y)$-directions, it is natural to let laser beams travel along these directions to create a standing wave pattern of intensity minima and maxima. In order to be able to control the strengths and phases of right and left circularly polarized light in the beams independently, however, the momentum vectors of the beams should have a nonzero $z$-component, which leads to the eight travelling directions $(\pm1,\pm1,\pm\beta)$ mentioned in the main text. We note that $\beta^2\approx2$ is a particularly convenient choice because the trapping lasers then have approximately the same frequency as the fields used for inducing hops and one can use the same set of excited states for implementing the trap and for implementing the hopping terms (the frequencies should, of course, not be exactly the same to avoid interference).

\begin{figure}
\includegraphics[width=0.75\textwidth]{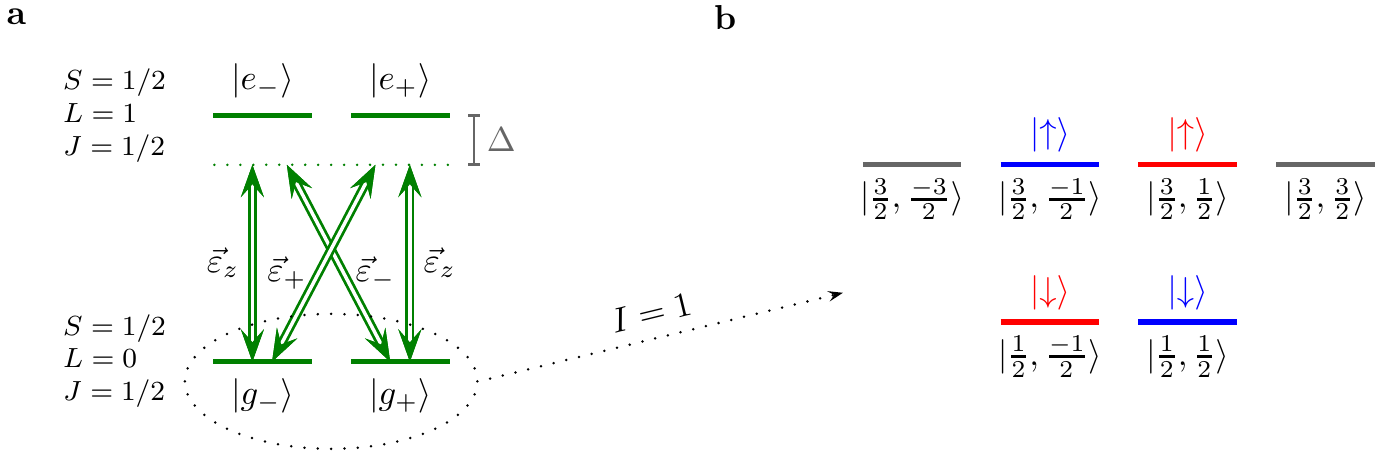}
\caption{\textbf{Atomic levels and light-atom interaction.} \textbf{a}, Off-resonant coupling of a $^2S_{1/2}$ ground state manifold to a $^2P_{1/2}$ excited state manifold with polarized light. $S$ is the spin angular momentum, $L$ is the orbital angular momentum, and $J$ is the total angular momentum obtained by coupling $S$ and $L$. \textbf{b}, Hyperfine levels of the ground state manifold when the nuclear spin is $I=1$. The states are labelled $|F,m_F\rangle$, where $F$ is the total angular momentum obtained by coupling $J$ and $I$, and $m_F$ is the projection of the total angular momentum on the $z$-axis. The choice of blue and red spin up and down states is indicated.}\label{FigS1}
\end{figure}

Let us demonstrate explicitly how this can be done for fermions with one valence electron, when the ground state manifold is a $^2S_{1/2}$ orbital, and the light fields couple the ground states off-resonantly to a $^2P_{1/2}$ orbital (see figure \ref{FigS1}a), which can be achieved with alkali atoms. We first create a standing wave pattern in the $(x+y)$-direction by adding up the fields
\begin{align}
\vec{E}_{1a}(\vec{r},T)
&=\textrm{Re}\left(E\left(\sqrt{2}\vec{\varepsilon}_+
+\alpha\sqrt{2}i\vec{\varepsilon}_-
+\frac{1}{\beta}(1+i)(1-\alpha)\vec{\varepsilon}_z \right)e^{ik_x(x+y)}e^{ik_zz}e^{-i\omega_1 T} \right),\\
\vec{E}_{1b}(\vec{r},T)
&=\textrm{Re}\left(E\left(\sqrt{2}\vec{\varepsilon}_+
-\alpha\sqrt{2}i\vec{\varepsilon}_-
-\frac{1}{\beta}(1+i)(1+\alpha)\vec{\varepsilon}_z \right) e^{-ik_x(x+y)}e^{ik_zz}e^{-i\omega_1 T}\right),
\end{align}
where $\vec{r}=(x,y,z)$, $T$ is time, $E=E^*$, $\vec{\varepsilon}_+=(-1,-i,0)/\sqrt{2}$ is the polarization vector of right circularly polarized light, $\vec{\varepsilon}_-=(1,-i,0)/\sqrt{2}$ is the polarization vector of left circularly polarized light, $\vec{\varepsilon}_z=(0,0,1)$ is the polarization vector of $z$-polarized light, $\alpha$ is a real parameter,
\begin{equation}
k_x=\frac{k}{\sqrt{2+\beta^2}}, \qquad k_z=\frac{k\beta}{\sqrt{2+\beta^2}}, \qquad k=\frac{2\pi}{\lambda},
\end{equation}
$\lambda$ is the wavelength of the fields, and $\omega_1$ is the frequency. Note that the intensity profile of the right (left) circularly polarized component is proportional to $\cos^2(k_x(x+y))$ ($\sin^2(k_x(x+y))$) and that the coefficients of $\vec{\varepsilon}_z$ have been chosen such that the polarizations of the fields are perpendicular to the directions of travelling as it should be for light fields. As seen in the figure, the presence of the $z$-polarized component causes the trapping lasers to induce Raman transitions between the two ground states. This undesired effect can be cancelled by adding the fields
\begin{align}
\vec{E}_{2a}(\vec{r},T)
&=\textrm{Re}\left(E\left(\sqrt{2}\vec{\varepsilon}_+
+\alpha\sqrt{2}i\vec{\varepsilon}_-
-\frac{1}{\beta}(1+i)(1-\alpha)\vec{\varepsilon}_z \right)e^{ik_x(x+y)}e^{-ik_zz}e^{-i\omega_2 T} \right),\\
\vec{E}_{2b}(\vec{r},T)
&=\textrm{Re}\left(E\left(\sqrt{2}\vec{\varepsilon}_+
-\alpha\sqrt{2}i\vec{\varepsilon}_-
+\frac{1}{\beta}(1+i)(1+\alpha)\vec{\varepsilon}_z \right) e^{-ik_x(x+y)}e^{-ik_zz}e^{-i\omega_2 T}\right),
\end{align}
where the frequency $\omega_2$ is assumed to be slightly different from $\omega_1$ such that the fields do not interfere coherently.

The fields
\begin{align}
\vec{E}_{3a}(\vec{r},T)
&=\textrm{Re}\left(E \left(\sqrt{2}\vec{\varepsilon}_+
+\alpha\sqrt{2}i\vec{\varepsilon}_-
-\frac{1}{\beta}(1-i)(1+\alpha)\vec{\varepsilon}_z \right)e^{-ik_x(x-y)}e^{ik_zz}e^{-i\omega_3 T} \right),\\
\vec{E}_{3b}(\vec{r},T)
&=\textrm{Re}\left(E\left(\sqrt{2}\vec{\varepsilon}_+
-\alpha\sqrt{2}i\vec{\varepsilon}_-
+\frac{1}{\beta}(1-i)(1-\alpha)\vec{\varepsilon}_z \right)e^{ik_x(x-y)}e^{ik_zz}e^{-i\omega_3 T}\right),\\
\vec{E}_{4a}(\vec{r},T)
&=\textrm{Re}\left(E \left(\sqrt{2}\vec{\varepsilon}_+
+\alpha\sqrt{2}i\vec{\varepsilon}_-
+\frac{1}{\beta}(1-i)(1+\alpha)\vec{\varepsilon}_z \right)e^{-ik_x(x-y)}e^{-ik_zz}e^{-i\omega_4 T} \right),\\
\vec{E}_{4b}(\vec{r},T)
&=\textrm{Re}\left(E\left(\sqrt{2}\vec{\varepsilon}_+
-\alpha\sqrt{2}i\vec{\varepsilon}_-
-\frac{1}{\beta}(1-i)(1-\alpha)\vec{\varepsilon}_z \right)e^{ik_x(x-y)}e^{-ik_zz}e^{-i\omega_4 T}\right),
\end{align}
do the same for the $(x-y)$-direction, where $\omega_3$ and $\omega_4$ are slightly different from $\omega_1$ and $\omega_2$ to avoid coherent interference.

Eliminating the excited states with a Schrieffer-Wolff transformation, we get the trapping potentials
\begin{align}
V_-&=-V_0(2\beta^2+\alpha^2-1)(\sin^2(k_x(x+y))+\sin^2(k_x(x-y)))+2V_0(2\beta^2+\alpha^2),\\
V_+&=-V_0(1+2\alpha^2\beta^2-\alpha^2)(\cos^2(k_x(x+y))+\cos^2(k_x(x-y)))+2V_0(1+2\alpha^2\beta^2),
\end{align}
for $|g_-\rangle$ and $|g_+\rangle$, respectively, where $V_0\propto E^2/(\beta^2\Delta)$ and the detuning $\Delta$, defined in the figure, is negative (positive) for red (blue) detuning. For $-V_0(2\beta^2+\alpha^2-1)>0$ and $-V_0(1+2\alpha^2\beta^2-\alpha^2)>0$, we thus conclude that $|g_-\rangle$ is trapped around the minima of $(\sin^2(k_x(x+y))+\sin^2(k_x(x-y)))$, whereas $|g_+\rangle$ is trapped around the minima of $(\cos^2(k_x(x+y))+\cos^2(k_x(x-y)))$, which gives the desired checkerboard pattern.

So far, we have considered the states at the level of fine structure. Assuming, as an example, that the nuclear spin is $I=1$, which is the case for $^6Li$, the ground state manifold in fact consists of six states as depicted in figure \ref{FigS1}b. The trapping potentials of these states are
\begin{align}
V_{\frac{3}{2},\frac{3}{2}}&=-V_0(1+2\alpha^2\beta^2-\alpha^2) (\cos^2(k_x(x+y))+\cos^2(k_x(x-y)))+2V_0(1+2\alpha^2\beta^2),\\
V_{\frac{3}{2},\frac{1}{2}}=V_{\frac{1}{2},-\frac{1}{2}} &=-V_0\left(\frac{4}{3}\alpha^2\beta^2 -\alpha^2+1-\frac{2}{3}\beta^2\right)\left(\cos^2(k_x(x+y))+\cos^2(k_x(x-y))\right) +2V_0\left(1+\frac{4}{3}\alpha^2\beta^2\right),\\
V_{\frac{3}{2},-\frac{1}{2}}=V_{\frac{1}{2},\frac{1}{2}} &=-V_0\left(\alpha^2-\frac{2}{3}\alpha^2\beta^2+\frac{4}{3}\beta^2-1\right) \left(\sin^2(k_x(x+y))+\sin^2(k_x(x-y))\right)
+2V_0\left(\alpha^2+\frac{4}{3}\beta^2\right),\\
V_{\frac{3}{2},-\frac{3}{2}}&=-V_0(\alpha^2+2\beta^2-1) (\sin^2(k_x(x+y))+\sin^2(k_x(x-y)))+2V_0(\alpha^2+2\beta^2).
\end{align}
For $-V_0\left(\frac{4}{3}\alpha^2\beta^2 -\alpha^2+1-\frac{2}{3}\beta^2\right)>0$ and $-V_0\left(\alpha^2-\frac{2}{3}\alpha^2\beta^2+\frac{4}{3}\beta^2-1\right)>0$, which for red detuning ($V_0<0$) amounts to
\begin{equation}
\frac{\frac{2}{3}\beta^2-1}{\frac{4}{3}\beta^2-1}<\alpha^2
<\frac{\frac{4}{3}\beta^2-1}{\frac{2}{3}\beta^2-1}, \qquad
(\textrm{or }1/5<\alpha^2<5\textrm{ for }\beta^2=2),
\end{equation}
we can get the desired potential by choosing the blue and red spin up and down states as shown in figure \ref{FigS1}b. Since the potentials shift the two blue states by the same amount and the two red states by the same amount, the energy difference between the blue and red spin up states is the same as the energy difference between the blue and red spin down states, which is what allows us to drive the hops of spin up and spin down with the same lasers. If this symmetry cannot be achieved in a given setup, however, one can double the number of standing wave laser fields used for implementing the hops and drive the transitions independently. The energies of the states $|\frac{3}{2},\frac{3}{2}\rangle$ and $|\frac{3}{2},-\frac{3}{2}\rangle$ are, in general, shifted by different amounts so that transitions to these states induced by the hopping lasers can be avoided automatically. Note also that the difference between the maxima and minima of the potentials can be adjusted. For $\beta^2=2$, e.g., the difference is $|-2V_0(-\frac{1}{3}\alpha^2+\frac{5}{3})|$ for the blue lattice potential and $|-2V_0(\frac{5}{3}\alpha^2-\frac{1}{3})|$ for the red lattice potential. Therefore $\alpha^2$ and $\beta^2$ allow us to control the ratio between the tunnelling rates between nearest-neighbour sites in the blue and red lattices. Finally, the energy of a fermion sitting on a blue lattice site must be different from the energy of a fermion sitting on a red lattice site to make the proposed implementation of the hopping terms work. If the contributions to this difference coming from the optical lattice potential and the field labelled b3 in the main text are not appropriate, one can control the difference by adding an additional laser field along the $z$-axis with right or left circular polarization.

\section*{Laser induced hopping terms}

The transitions induced by the laser fields r1, r2, and b3 in the main text can be understood by noting that the transition amplitudes are proportional to the spatial integral of the product of the Wannier functions at the two sites and the two field components that are involved in the transition. Since the Wannier functions decay rapidly, there are only transitions between nearby sites.

Let us first consider the jumps between neighbouring blue and red lattice sites induced by r1 and b3. The transition amplitude for jumps along the $y$-axis vanishes because the Wannier functions and b3 are even functions and r1 is an odd function with respect to reverting the $x$-axis around the $x$-coordinate of the sites. Jumps along the $x$-axis are allowed, and since the sign of the amplitude of r1 between two neighbouring sites is alternatingly plus and minus, the signs of the hopping amplitudes also alternate. r2 and b3 similarly induce hops in the $y$-direction, but not in the $x$-direction. For the choice of hyperfine levels considered above, it is the right (left) circularly polarized component of b3 that is involved when a fermion in the spin up (down) state jumps. Therefore one can make the hopping amplitudes for up and down spins equal by adjusting $E_\pm$ appropriately, i.e., by adjusting the polarization state of b3.

Let us also consider transitions involving one photon from r1 and one photon from r2. In this case, the fermion cannot change its internal state due to energy conservation, and it cannot jump to a site of different colour. If the fermion remains at the same site, the transition amplitude is zero due to symmetry. If the transition involves a jump to a next-nearest neighbour site, there is destructive interference between absorbing a photon from r1 and emitting a photon into r2 or absorbing a photon from r2 and emitting a photon into r1, and therefore no transitions occur.

r1, r2, and b3 also induce transitions, where a fermion absorbs and emits a photon from the same beam without changing its internal state and either remains at the same site or jumps to a next-nearest neighbour site. The former case gives rise to the trapping in the $z$-direction and also slightly modifies the potential in the $xy$-plane. The modification can, however, be made insignificant by decreasing $E$ and increasing $E_\pm$ without affecting the nearest-neighbour hopping amplitudes that are proportional to $EE_\pm$. The latter case gives rise to a contribution to the next-nearest neighbour hopping terms in the Hamiltonian. For a given choice of $E_\pm$ it may happen that hops from blue to blue lattice sites occur at a different rate than hops from red to red lattice sites. As mentioned above, however, one can lower the potential barriers between the sites for the sublattice with slow jumps relative to the other to compensate.

\end{document}